\documentclass[amsmath,amssymb,aps,nofootinbib]{revtex4}
\usepackage[dvips]{graphicx}
\usepackage{graphicx}
\usepackage{amsfonts}
\usepackage{bm}
\usepackage{amsmath}
\usepackage{amssymb}
\usepackage{color}
\usepackage[all]{xy}

\begin{document}

\title{Flux Quantization in Dilatonic Taub--NUT Dyons}
\author{Daniel \surname{Flores-Alfonso}$^{1,}$}
\email[]{daniel.flores@correo.nucleares.unam.mx}
\author{Hernando \surname{Quevedo}$^{1,2,}$}
\email[]{quevedo@nucleares.unam.mx}
\affiliation{$^1$Instituto de Ciencias Nucleares,
Universidad Nacional Aut\'onoma de M\'exico,\\
AP 70543, Ciudad de M\'exico 04510, Mexico \\
$^2$Institute of Experimental and Theoretical Physics,
Al-Farabi Kazakh National University, 
Almaty 050040, Kazakhstan}

\begin{abstract}
Spacetimes that include a boundary at infinity have a
non-trivial topology. 
The homology of the background influences
gauge fields living on them
and lead to topological charges.
We investigate the charges and fluxes of 
fields over a Taub--NUT background in 
Einstein--Maxwell dilaton--axion gravity,
by using the relative homology and 
de Rham cohomology. 
It turns out that the electromagnetic 
sector is devoid of restrictions from a topological viewpoint.
There are, however, flux quanta for the axion and dilaton fields.
These results are obtained from
the absolute homology of the spacetime boundary.
The solutions we probe originate in
the four dimensional low energy limit of heterotic string theory.
So our results are complemented by
the stringy coupling present in the fields.
The quantization has a bundle theoretic interpretation as the axion's flux
corresponds to the topological index of an underlying 2-bundle.
\\[0.2cm]
\end{abstract}

\maketitle

\section{Introduction}

Topological structures are responsible for a variety of theorized and observed
quantum phenomena. In superconductors, for instance,
magnetic flux quantization is a topological feature \cite{thouless}.
SQUID magnetometers are used to measure the flux of superconducting 
rings. This results in a magnetic flux that can only take integer multiples
of $h/2e$. More precisely, the number of flux quanta is related to the winding number 
of a condensate wave function.

In certain quantum, or even semi-classical, theories, topological quantization of 
the type described above is closely related to wave functions and their behavior.
A close example is string theory, where topological structures and compactification
lead to quantized momenta and charge from winding numbers of strings \cite{dbranes}.
In gravity theories, this has not been done, as of yet. 

On the other hand, Kaluza--Klein compactification explores the idea that
four dimensional observers could live
in higher dimensions, but that extra dimensions are directly inaccessible.
Let $M=\mathbb{R}^4\times S^1$
be a five-dimensional space with a compact direction along a circle.
Einstein gravity in $M$ is perceived
as Einstein--Maxwell dilaton gravity by observers in $\mathbb{R}^4$.
Massless fermions on $M$ are seen as
massive, charged fermions in the four dimensional viewpoint.
Moreover, the electric charge becomes 
quantized as an effect of the winding of a field along
the circular dimension \cite{kk}.

String theory has generalized this approach as physical strings are supposed
to exist in spacetime dimensions higher than four. 
Moreover, an important distinction arises, namely,
the presence of fields produced by ``electrically'' charged
strings. Just as point-like particles generate
Abelian one-form gauge potentials, so  strings 
generate Abelian two-form potentials.
Therefore, the field strength symmetry is
\begin{equation}
 H_{3}\rightarrow H_{3}+{\rm d} B_{2}.
\end{equation}
Having an exact solution for the full classical action of string theory
is the best of the scenarios; however, this is not possible yet. 
The next to best case is having an exact solution to the low energy classical action.
But starting from a four dimensional
low energy string solution does not allow for a quantization
following from a compactification scenario, as mentioned above.
Nevertheless, topological methods have been formulated to study
Abelian fields over general spacetimes \cite{ao}.

Motivated by understanding topological quantum 
numbers in gravitation, in this work we study
the field fluxes of the bosonic part of a truncation 
four dimensional low energy
heterotic string theory and the homology of
the spacetime on which they gravitate.
Three of these fluxes come from the Abelian gauge fields.
The truncated effective low energy string action can be rewritten 
as an Einstein--Maxwell dilaton-axion theory.
However, the stringy context from where they 
originate leaves a special coupling between
the fields.
A particular family of solutions to which our
discussion applies is that of
Taub--NUT-like  backgrounds \cite{taub,nut,misner}. These solutions
were independently discovered in \cite{jm} and \cite{ss}.
 The methods we will consider in this work apply to any
theory where gauge fields self-gravitate;
notwithstanding, we report on physical interpretations
that only follow from the stringy coupling.

Spacetimes that include a boundary allow for 
higher-order Abelian gauge field charges.
The role the boundary plays is that of a place
for the electric lines of force to land on.
This framework was originally developed for brane
sources which are electrically charged.
The distinction between charges and fluxes is emphasized,
as they relate to cycles of different homological type.
By using the
magnetic flux conditions of extended objects \cite{teitelboim},
a Dirac quantization condition \cite{dirac} is proved for
branes of electric type of arbitrary order, living in
spacetimes of any dimension \cite{ao}. 
This Dirac condition relates the charge created
by an electric current and the magnetic flux created by the field. 

We write the truncated four dimensional low energy
effective heterotic string action in the Einstein 
frame\footnote{As opposed to the string frame that
arises naturally in string theory.} as
\begin{equation}
 \int {\rm d}^4x \sqrt{-g}\left(R-\frac{1}{2}\left(\nabla\phi\right)^2-
 \frac{1}{12}e^{-2\phi} H^2-\frac{1}{8}e^{-\phi} F^2\right).
\label{action}
\end{equation}
The background metric $g$ is coupled to three matter 
fields: a scalar field $\phi$, the dilaton, and
two Abelian gauge fields $F$ and $H$.
The electromagnetic field strength $F$ is locally represented by
a two-form $F$. The field $H$ is a higher gauge field
and is locally represented by a three-form. These field strengths are determined by
Abelian gauge potentials $A$ and $B$ and comply with
\begin{subequations}
\begin{align}
 F&={\rm d}A,
 \label{fda}\\ 
 H&={\rm d}B-\frac{1}{4}A\wedge F.
 \label{ax}
\end{align} 
\end{subequations}
Notice how the Maxwell sector contributes to the
$H$ field through a Chern--Simons term. It is said that this Chern--Simons term
twists the Kalb--Ramond two-form $B$.

Variation of the effective string action (\ref{action}) yields the Einstein field equations
\begin{align}
 G_{\mu\nu}&=
 \frac{1}{2}\left(\partial_{\mu}\phi\partial_{\nu}\phi-
 \frac{1}{2}g_{\mu\nu}\left(\nabla\phi\right)^2\right)
 +\frac{e^{-\phi}}{8}\left(2F_{\mu\alpha}F_{\nu}^{~\alpha}
 -\frac{1}{2}g_{\mu\nu}F^2\right)\notag\\
 &+\frac{e^{-2\phi}}{12}\left(3H_{\mu\alpha\beta}H_{\nu}^{~\alpha\beta}
 -\frac{1}{2}g_{\mu\nu}H^2\right),
\end{align}
as well as the equations of motion for the dilaton field
\begin{equation}
 {\rm d}(\star {\rm d}\phi)=-\frac{e^{-\phi}}{4}F\wedge \star F-
 e^{-2\phi}H\wedge \star H,
 \label{phi}
\end{equation}
which is written in this way for later
convenience. The Maxwell equation is
\begin{equation}
 {\rm d}\star (e^{-\phi}F)= \star e^{-2\phi}H\wedge F,
 \label{f}
\end{equation}
while the $H$ field obeys
\begin{equation}
 {\rm d}\star (e^{-2\phi}H)=0.
 \label{h} 
\end{equation}

In the following section, we introduce the concept of 
flux and distinguish it from charge.
We comment on how the fluxes of the Abelian fields 
we investigate are related to circular
bundles and two-bundles. The Maxwell field leads 
to two fluxes electric and magnetic.
The higher gauge field $H$ determines a flux as well,
which by analogy is of magnetic type.
Equation (\ref{h}) shows that $H$ can be effectively 
replaced by a dual scalar field. This field
is the axion. Comparing the axion and dilaton allow for
a fourth flux to be defined which is of electric type.

The backgrounds we consider share the global
structure of the Taub--NUT vacuum solution,
as well as many of its local similarities. 
These dilatonic Taub--NUT backgrounds are 
described in detail in section \ref{nuts}
In section \ref{fluxq} we present the 
flux quantization which arises in these configurations 
due to nontrivial topological structures
which underline them. Finally, in section \ref{disc}
we summarize our results and compare them with previous
research. We also comment on how the solutions we have studied
fit into a larger class of solutions.
Some of our results are particular to the special cases presented
here and some are valid in general.

\section{Charges and Fluxes}

In this section, we introduce the concepts of charge and flux
that apply to differential forms of any degree, 
representing field strengths and currents.
We are interested in describing fluxes which resemble those of
Maxwell theory. Maxwell's equations are statements about
 a differential two-form $F$, which represents the field strength. 
The gauge symmetry of these fields can be encoded in U(1)-bundles,
and the study of their topological invariants has led to
many interesting physical effects. In string theory, similar fields
arise which are also Abelian in nature, i.e.,
\begin{equation}
 C_{n}\rightarrow C_{n}+{\rm d}B_{n-1},
\end{equation}
however, they lead to more complicated bundle structures, $n$-bundles.
In a principal $n$-bundle, the fiber space corresponds to an $n$-group.
Since the U(1) structure can be restated in terms of spacetime 
cohomology,  a similar approach can be performed for field strengths
of higher rank. This is the main idea of our method and for which we
follow the concepts presented in \cite{ao}. Therein,
the distinction is stressed between charges and 
fluxes as they are classified by distinct homological structures of spacetime. 
Charges are produced by currents and fluxes are produced by fields. Of course,
both are related but we will remark their differences in the following analysis. 

This work focuses on a dilatonic Taub--NUT space \cite{jm} that arises
from a truncation of four dimensional low-energy effective heterotic string theory. 
The solution has a scalar dilaton field, 
an electromagnetic field, and a Kalb--Ramond-like field.
Since currents are physically absent we focus only on fluxes.
The equations of motion are more complicated than in the case of
Maxwell's theory, because of string theory intricacies,
but they allow for an appropriate interpretation
for these topological methods to hold.

To distinguish charges from fluxes, let us first consider the action
\begin{equation}
 \int\limits_{M} e^{-\phi}F\wedge \star F+A\wedge \star j,
\end{equation}
defined over a spacetime $M$. This action together with $F={\rm d}A$ yield
\begin{subequations}
\begin{align}
 {\rm d}F&=0,\label{maxwell0}\\ 
 {\rm d}\star(e^{-\phi}F)&=\star j,\label{maxwell}
\end{align} 
\end{subequations}
which are Maxwell's equations in the presence of a dilaton field $\phi$. 
We can immediately conclude that $\star j$ is a conserved current. However,
physical currents are also localized. So we have both
\begin{subequations}
\begin{align}
 {\rm d}\star j&=0,\label{dstarj}\\ 
 \star j|_{\partial M}&=0.\label{starj}
\end{align} 
\end{subequations}
Here, we have in mind a manifold which includes a spacetime and its asymptotic boundary.
In other words, Eq.\ (\ref{starj}) is a boundary condition. This condition may be relaxed
in a compact manifold; however, the analysis of this case is beyond the scope of the present work.
The charge on a (hyper)surface $S\subset M$ is defined by
\begin{equation}
 Q_E(S)=\int\limits_{S}\star j,
\end{equation}
for $\partial S\subset \partial M$. If we now consider $C\subset M$ and
$\beta\subset \partial M$, then conditions (\ref{dstarj}-\ref{starj}) imply
\begin{equation}
 Q_E(S+\partial C+\beta)-Q_E(S)=
 \int\limits_{\partial C}\star j+\int\limits_{\beta}\star j=
 \int\limits_{C}{\rm d}\star j=0,\label{qe}
\end{equation}
where we have used Stokes' theorem.
We conclude that the electric charge is an invariant of the
relative homology of spacetime modulo its boundary.

As is well-known, the nontrivial geometry of spacetime in electrovacuum
can lead to solutions without current, but with a nontrivial
field strength. This is an indication of the existence of a field flux
without charge. Electric flux is a simpler object than electric charge;
it is defined only 
through the field without reference to currents, i.e.,
\begin{equation}
 \Phi_E(\beta)=\int\limits_{\beta}\star F,
\end{equation}
which is is an invariant of the absolute homology of the
spacetime boundary. Likewise, magnetic flux is linked to the
absolute homology of spacetime itself according to the definition
\begin{equation}
 \Phi_M(b)=\int\limits_{b} F,
\end{equation}
where $\partial b=0$. Of course, charges do lead to fluxes as
can be concluded from Eq.\ (\ref{maxwell}), i.e., 
\begin{equation}
 Q_E(S)=\int\limits_{S}\star j=
 \int\limits_{S}{\rm d}\star (e^{-\phi}F)=
 \int\limits_{\partial S}\star F,
 \label{chf}
\end{equation}
whenever $\phi|_{\partial M}=0$. In Eq.\ (\ref{chf}) 
we used Stokes' theorem just as we did in relation (\ref{qe}).
Once more we emphasize that we are interested here
only in electric and magnetic type fluxes. Currents will be absent
and hence charges are null.

Finally, we mention that the equations of motion that
interest us intermingle the fields, but they can be written in a way
closely following Maxwell's Eqs.\ (\ref{maxwell0}-\ref{maxwell}).
The role of the currents are effectively played by the
field content that behaves as in Eqs.\ (\ref{dstarj})--(\ref{starj}).

\section{NUT Spaces from String Theory}
\label{nuts}

The first dilatonic Taub--NUT space that was found came
from studying supersymmetric conformal field theories \cite{johnson}.
The exact field theory solution describes an extremal
Taub--NUT throat geometry, in the same sense two dimensional black holes
are described by conformal field theories \cite{wittencft}.
The exact solution to the low energy heterotic string
theory was found shortly after \cite{jm} and has the following geometry 
\begin{equation}
{\rm d}s^2 = -\frac{f_1}{f_2}({\rm d}t+(x+1)l{\rm d}\varphi)^2+
\frac{f_2}{f_1}{\rm d}r^2+f_1(r^2+l^2){\rm d}\Omega^2,
\label{g}
\end{equation}
where ${\rm d}\Omega^2$ stands for the round metric on the
unit two-sphere coordinated by ($\theta$, $\varphi$). The metric functions
$f_1(r)$ and $f_2(r)$ are given by
\begin{subequations}
 \begin{align}
 f_1&=1-2\frac{mr+l^2}{r^2+l^2},\\ 
 f_2&=1+(x-1)\frac{mr+l^2}{r^2+l^2}.
\end{align}
\end{subequations}
The parameters $m$, $l$ and $x$ are related to the mass
$M$ and the NUT parameter $N$ of the background. 
Examining the asymptotic behavior of the metric we see that
\begin{subequations}
 \begin{align}
 M&=\frac{(x+1)m}{2},\label{mass}\\
 N&=\frac{(x+1)l}{2}.\label{nut}
\end{align}
\end{subequations}
The rest of the solution is given by the fields
\begin{align}
 \phi&=-\ln{f_2},
 \label{ln}\\
 B&=
 \frac{f_1}{f_2}(x-1)l\cos\theta {\rm d}t\wedge {\rm d}\varphi,\\
 A&=
 \frac{\sqrt{x^2-1}}{f_2}
 \left[(1-f_1){\rm d}t-2l\cos\theta f_1 {\rm d}\varphi\right].
 \label{a}
\end{align}
This solution is generated by boosting a Taub--NUT space
with constant dilaton field. Let us recall that
this is a trivial solution to the string motion;
nonetheless, it can be used to generate new solutions,
by virtue of the O(1,1) symmetry of the equations of motion. 
Indeed, as we sill see in the following section,
new solutions are generated when considering the
full symmetries of the motion 
equations.

The NUT parameter is sometimes called magnetic mass.
Equations (\ref{mass}) and (\ref{nut}) bear resemblance
to the electric and magnetic parameters of Maxwell theory.
The asymptotic behavior of the Maxwell field yield
\begin{subequations}
 \begin{align}
 Q_E&=2m\sqrt{x^2-1},\\
 Q_M&=2l\sqrt{x^2-1}.
\end{align}
\end{subequations}

A similar approach with the Kalb--Ramond field strength yields
\begin{equation}
Q_{KB}=l(x-1).\label{qkb}
\end{equation}

Now, using Eq.\ (\ref{h}), we can define the scalar field $a$ by
\begin{equation}
 H = -e^{2\phi}\star {\rm d}a,
\end{equation}
which together with Eqs.\ (\ref{ax}) and (\ref{f}) portray the
motion as Maxwellian in the presence of an axion field. 
Similar to the definition of the ``magnetic charge'' $Q_M$
given above, we also define the axion ``charge'' $Q_A$ in terms of the $H$ flux.
Since the Hodge dual $\star H$ vanishes at the boundary,
then the $H$ field only produces a magnetic type of flux. 
This means that the
axion field produces only a flux of electric type.
From Eq.\ (\ref{h}) we see that the solution under consideration yields
\begin{equation}
 {\rm d}\left(H|_{\partial M}\right)=0,
\end{equation}
indicating the existence of an additional conserved flux.
As of yet, we have not discussed Maxwell's equations with magnetic type currents, 
so we simply rewrite the equations for $H$ in an electric fashion
\begin{equation}
 {\rm d}\star(e^{2\phi} {\rm d}a)=\frac{1}{4}F\wedge F,\label{ax2}
\end{equation}
where we have taken the exterior derivative of
equation (\ref{ax}). Thence the axion's parameter is defined asymptotically
as well, coinciding with the result in Eq.\ (\ref{qkb}).

However, we may also use the following integral definition
\begin{equation}
 Q_A=\frac{1}{4\pi\beta}\int\limits_{S^3_\infty} \star {\rm d}a=l(x-1).
\end{equation}
Here $\beta=8\pi N=4\pi(x+1)l$ is the period of the time-like
coordinate $t$.
This definition is motivated by the way electromagnetic fluxes are
defined through integrals in asymptotically flat (AF) spacetimes, e.g.,
\begin{equation}
 Q_E=\frac{1}{4\pi}\int\limits_{S^2_\infty} \star F\quad\text{(in an AF spacetime)}.
\end{equation}
This parameter is not independent from the ones presented before.
Moreover, in a similar way as above we write
\begin{equation}
 Q_D=\frac{1}{4\pi\beta}\int\limits_{S^3_\infty} \star {\rm d}\phi=m(x-1),
\end{equation}
for the dilaton's parameter. As a final comment we 
mention that this solution is the Taub--NUT generalization
of the unique static black hole 
in dilaton Einstein--Maxwell theory. In the black hole solution
the dilaton's ``charge'' is not exactly a hairy 
parameter as it is not independent of the black hole's mass
and electromagnetic parameter. Here the same interpretation holds.

\subsection{The SL(2,$\mathbb{R}$) Symmetry in the Motion}
\label{symm}

The solution described in the previous section serves
as a seed to generated new solutions. The generating technique is
based on the SL(2,$\mathbb{R}$) symmetry that characterize
the string equations of motion in four dimensions.
The symmetry group is generated by $S$ duality and $T$ duality
transformations that we describe below. In this context, it is convenient 
to define
the complex scalar $\lambda$ and complex two-forms $F_{\pm}$ by
\begin{subequations}
 \begin{align}
  \lambda&=a+{\rm i}e^{-\phi},\\
  F_{\pm}&=F\pm \star {\rm i}F.
 \end{align}
\end{subequations}
Transformations of type $T$ shift
$\lambda$ by a real constant $y$,
\begin{equation}
 T: \lambda\rightarrow \lambda+y,
\end{equation}
leaving $F_+$ and $F_-$ fixed.
Transformations of type $S$ are more complicated but
are presented concisely as
\begin{equation}
 S: (\lambda,F_+,F_-)\rightarrow
 (-\frac{1}{\lambda},-\lambda F_+,-\lambda^* F_-),
\end{equation}
where $\lambda^*$ is the complex conjugate of $\lambda$.
The $S$-transform of the solution
(\ref{g}, \ref{ln}-\ref{a}) corresponds to a stringy
electromagnetic dual, where  
the electric and magnetic parameters are interchanged in the usual manner. 
The background geometry is invariant and the $S$-dual fields are given by
\begin{align}
 \hat{\phi}&=
 -\ln{\left\{\frac{(r^2+l^2)f_2}{[r+(x-1)m]^2+x^2l^2}\right\}},\\
 \hat{B}&=
 -\frac{(r-m)^2+x(m^2+l^2)}{(r^2+l^2)f_2}(x-1)l\cos\theta
 {\rm d}t\wedge {\rm d}\varphi,\\
 \hat{A}&=
 \frac{\sqrt{x^2-1}}{f_2}\left\{\frac{2l(r-m)}{(r^2+l^2)}{\rm d}t\right.
 \notag\\
 &+\left. 2\cos\theta\frac{mr^2+[(x-1)m^2+(x+1)l^2]r-ml^2}{(r^2+l^2)}
 {\rm d}\varphi\right\}.
\end{align}
Although the fields themselves are modified,
the fluxes they produce remain quite similar, i.e., 
 the axionic and dilatonic fluxes only change by a
 sign and the electric/magnetic 
duality is present in the U(1) fluxes.
The changes in the parameters can be symbolically summarized as
\begin{align}
 \hat{Q}_E&=Q_M, & \hat{Q}_M&=-Q_E,\\
 \hat{Q}_A&=-Q_A, & \hat{Q}_D&=-Q_D.
\end{align}

Furthermore, a one-parameter family of solutions can be
generated applying $S$ and $T$ transformations \cite{twist}, maintaining as before
the geometry unaffected. The fields now depend on a parameter $y$ such that $y=\infty$
yields the original solution (\ref{ln})-(\ref{a}) and
$y=0$ gives its $S$-dual. These solutions are given by 
\begin{align}
 \tilde{\phi}&=
 -\ln{\left[\frac{(y^2+1)f_2}{f_2^2+f_3^2}\right]},\\
 \tilde{B}&=
 -\frac{(x-1)}{(y^2+1)(r^2+l^2)f_2}[(l+2ym-y^2l)(r-m)^2\notag\\
 &+(xl-2ym+y^2l+(x+1)yr)(m^2+l^2)] \cos\theta {\rm d}t\wedge {\rm d}\varphi,\\
 \tilde{A}&=
 \frac{2\sqrt{x^2-1}}{\sqrt{y^2+1}}\left\{f_4 {\rm d}t+[m-yl+(x+1)lf_4]\cos\theta
 {\rm d}\varphi\right\}.
\end{align}
where we have defined
\begin{subequations}
 \begin{align}
 f_3&=(x-1)l\frac{r-m}{r^2+l^2}-y,\\
 f_4&=\frac{l(r-m)+y(mr+l^2)}{(r^2+l^2)f_2}.
 \end{align}
\end{subequations}

These solutions correspond to an interpolation
between the extreme values $y=0$ and $y=\infty$.
A trigonometric reparametrization of $y=\tan\Theta$ allows us to write
\begin{equation}
 \tilde{Q}_E=Q_E\sin\Theta+Q_M\cos\Theta,
 \qquad\tilde{Q}_M=Q_M\sin\Theta-Q_E\cos\Theta,
 \label{qm}
\end{equation}
for the electric and magnetic parameters of the
interpolation solutions. A similar pair of equations follows for the
``charges'' of the dilaton and axion scalar fields given by
\begin{equation}
 \tilde{Q}_D=-Q_D\cos2\Theta-Q_A\sin2\Theta,
 \qquad\tilde{Q}_A=-Q_A\cos2\Theta+Q_D\sin2\Theta.
 \label{qs}
\end{equation}
So for all values of $\Theta$,
the quantity $\tilde{Q}_E^2+\tilde{Q}_M^2$ is preserved, which is expected of
a dyonic sector. The dilaton--axion sector follows
this behavior as well, signaling the term ``triply dyonic''
for these solutions; recall that the gravity
sector is dyonic because of the NUT charge.

Turning off the NUT parameter ($l=0$) leaves the
renowned family of dilaton black holes studied in \cite{g,gm,ghs}.
In general, all four fields remain nontrivial and
there are still four conserved fluxes.
The natural topology has now changed to what is expected for a black hole. 
The homology of the boundary $S^1\times S^2$,
however, remains non-trivial. Moreover,
the geometry has now gained nonretractable two-cycles,
which means that there are now topological U(1) fluxes
whose quanta correspond to winding numbers.
We will redefine the dilatonic flux (per unit solid angle) in the following manner
\begin{equation}
-\frac{1}{4\pi}\int\limits_{S^2_{\infty}} \iota_X (\star {\rm d}\phi),
\end{equation}
where $\iota_X$ is the interior product with respect to $X=\partial/\partial t$,
the timelike direction,
and the flux generated by the axion is redefined in a similar fashion. 
The entire flux content of the black hole 
is consistent with Eqs.\ (\ref{qm})-(\ref{qs})
if $Q_M=0=Q_A$ which, of course, follows from the condition
that $l=0$. Notice that the solutions at
$\Theta=0,\pi/2$ lack flux from the axion and the U(1) flux is either
electric or magnetic. For these solutions,
the absence of a dyonic background implies absence of dyonic fluxes.
In other words, they are either triply dyonic or not at all. 

Taub--NUT-like spacetimes are characterized by being dyonic,
but also for having closed timelike curves.
Notwithstanding, their inner horizon region is causal
and is Gowdy $S^3$ symmetric \cite{gowdy,moncrief,bh}.
This is comparable to how the inner horizon of Schwarzschild spacetime 
is equivalent to a Kantowski--Sachs spacetime \cite{ks}
or, more generally,
to how the inner horizon region
of the Kerr black hole corresponds to an unpolarized
Gowdy $S^1\times S^2$ spacetime \cite{hor}. 
When the inner region of the Kerr black hole
is made Gowdy symmetric,  then necessarily the exterior has
closed timelike curves. However, they are not dyonic.

\section{Flux Quantization}
\label{fluxq}

Having described the details of the configurations of 
interest we now continue to investigate how
the field fluxes we have mentioned must comply with certain 
quantization conditions. The origin
of these conditions is topological. For concreteness we focus 
mainly on the first solution presented, Eqs.\ (\ref{g}-\ref{a}),
and then comment related ones.

\subsection{The Maxwell Sector}

The methods described in Ref.\ \cite{ao} apply to
differential forms of distinct degrees such as those
presented in Eqs.\ (\ref{phi})-(\ref{h}).
We begin by applying those methods to the familiar Maxwell
equations. First, let us notice that on the boundary $\partial M$ 
we have both
\begin{equation}
 F|_{\partial M}\neq 0\qquad{\rm and}\qquad\star F|_{\partial M}\neq 0,
\end{equation}
meaning, in principal, that the Maxwell field produces a magnetic flux as well as 
an electric one. 
Equations (\ref{fda}) and (\ref{f}) together with the solution (\ref{g}-\ref{a})
imply
\begin{equation}
 {\rm d}\left(F|_{\partial M}\right)=0,
 \qquad {\rm d}\left(\star F|_{\partial M}\right)=0.
\end{equation}
So both electric and magnetic fluxes are conserved quantities.
Recall the magnetic flux is
\begin{equation}
 \Phi_M(S^2_{\infty})=\int\limits_{S^2_{\infty}}F,
 \label{mflux}
\end{equation}
and a similar formula holds for the electric counterpart.

These conserved fluxes suggest looking into the
homologies $H_2(M)$ and $H_{4-2}(\partial M)$. 
The spacetime topology is $\mathbb{R}\times S^3$ with boundaries
at $r=\pm\infty$ which are three-spheres.
Instead of considering a boundary $S^3\sqcup S^3$, 
we will only consider the one at $r=\infty$ for 
the sake of simplicity. There is no loss of
generality by applying this procedure; in fact, 
this has been done already in Eq.\ (\ref{mflux}). 
Since $H_2(S^3)=0$, the topology does not hint at 
there being a quantized magnetic flux or that its associated
magnetic charge $Q_M$ is topological. In fact, both
the electric and magnetic flux vanish since the two-sphere at
infinity is homologous to a point in the boundary.

Additionally, Eq.\ (\ref{f}) can be rewritten as
\begin{equation}
 {\rm d}\left(e^{-\phi}\star F+aF\right)={\rm d}\bar{F}=0,
 \label{barf}
\end{equation}
which is a magnetic type equation for $\bar{F}$ and is
classified as well by $H_2(M)$.
As both scalar fields vanish on the boundary,
the magnetic type flux of $\bar{F}$ corresponds to the
electric flux of $F$. 

These results stand out at first since most encounters
with magnetic monopoles and the like result in some form of
topological quantization. NUT  spaces are exceptions
to this criteria. The words ``counterexample to almost anything''
come to mind \cite{misnerbook}.

We close this section commenting on the relation that
these asymptotic fluxes have with
the topological indices of the underlying 
circle bundles preset. Notice that
$F$($\bar{F}$) is a closed form on $M$ and, in fact, it yields
the curvature of the U(1)-bundle
over $M$. At the boundary, there is a subbundle 
over $S^2_{\infty}$ whose curvature is given 
by $F$($\star F$) restricted to this submanifold.
The Chern number of this subbundle is the magnetic (electric) flux,
up to a scaling factor.
Moreover, there is another subbundle of interest,
the boundary itself.
There $\star F$ is closed and a circle bundle
may be constructed  as well. This construction connects
the group $H_2(\partial M)$ with the classification of the
principal U(1) bundles that can be constructed with the 
curvature $\star F$. In other words, the classification
scheme of Ref.\ \cite{ao} is consistent with the bundle
theoretic approach.

\subsection{Axions and Dilatons}
\label{axidil}

We start this section examining the higher gauge field $H$. 
As we saw above the effective four dimensional theory
we focus on is subject to a nonlinear
Kaluza--Klein effect stemming from heterotic string theory.
This ultimately couples the two Abelian sectors
of the effective theory. Specifically,
it twists the $B$ potential with a Chern--Simons 
term from the Maxwell sector [see Eq.\ (\ref{ax})].
Solution (\ref{g}-\ref{a}) shows that the $H$ flux is conserved.
Note that the Chern--Simons term on the right-hand side
acts effectively as a current, i.e.,
it satisfies Eqs.\ (\ref{dstarj})-(\ref{starj}).
The relevant homology group is $H_3(\partial M)=H_3(S^3)=\mathbb{Z}$.
Hence, the flux is topologically quantized
\begin{equation}
 \int\limits_{S^3_\infty}H=4\pi\beta Q_A=\varPhi_0 n.
 \label{aflux}
\end{equation}
where $\varPhi_0$ is the axion flux quantum and $n$ is some integer.
It is therefore clear that the flux may only
take discrete values. Chern numbers
have a standard normalization which leads to integer values
when integrating over spheres. The stringy couplings of the solution 
[see Eq.\ (\ref{ax2})] suggest
dividing by $\pi^2$ so that relation (\ref{aflux}) resembles a
Chern--Simons integral with standard normalization for U(1) fields. 
The actual Chern--Simons term in $H$
vanishes at the boundary, notwithstanding,
the flux integral does not.
Let us write this through the dual incarnation of the higher gauge field, the axion,
\begin{equation}
 \frac{1}{\pi^2}\int\limits_{S^3_\infty}\star {\rm d}a=4Q_M^2=
 \text{integer}.
 \label{pi}\tag{\ref{aflux}'}
\end{equation}

Therefore, the topological quantization of the axionic flux
quite ironically leads to the quantization of the magnetic parameter $Q_M$.
In spite of there being no homological indications in the magnetic Maxwell sector
of a quantization condition and even though the magnetic flux is zero.
The square of the magnetic parameter
takes integer values but not necessarily $Q_M$ itself.
This should be contrasted to situations where the 
Chern number represents a squared integer (see, e.g., \cite{dfhq}).

From the point of view of a higher gauge theory,
the Kalb--Ramond field lives on a circular
2-bundle \cite{baez}. The field is twisted and
so is classified by a twisted cohomology.
However, there is a sub-2-bundle at the boundary
where the field untwists.
The asymptotic Kalb-Ramond field is a representative
of ordinary cohomology.
The axion flux represents a topological number
associated to this sub-2-bundle.
This parallels the relation between Chern number and
magnetic flux in the subbundle mentioned in the previous section.

There is still one more field that can be discussed here. 
We will take ${\rm d}\phi$ as the form representing the field
strength of the dilaton.
The field leads to a purely electric flux ---
as the dilaton $\phi$ vanishes on the boundary. 
Once again we see [from Eq.\ (\ref{phi})] that
our solution of interest has a conserved flux.
Although the dilaton is not an Abelian gauge field, in the particular
solution we study it parallels the axion and so has an associated
effective flux
\begin{equation}
 \frac{1}{\pi^2}\int\limits_{S^3_\infty}\star {\rm d}\phi=4Q_EQ_M.
\end{equation}
However, due to the axion's flux quantization above we have
that the dilatonic flux at infinity is
\begin{equation}
 \Phi_D(S^3_\infty)=n\pi^2Q_E/Q_M,
\end{equation}
for the same integer $n$ in equation (\ref{aflux}).

\section{Discussion}
\label{disc}

We have studied dilaton--axion
fields in a truncated four dimensional low energy effective
heterotic string theory, particularly, on dyonic backgrounds. 
Our methods apply to a class of Taub--NUT-like spacetimes with
dyonic behavior in the Maxwell and scalar sectors.
We find that by the nature of the axion field and how it relates
to the Abelian fields,
the notions of electric or magnetic are quite loose for 
the U(1) and axion sectors.
This means that electric and axionic flux can consistently be
portrayed both as electric or magnetic by virtue 
of the nature of the equations of motion. 
This class of triply dyonic solutions have a quantized flux for
both the axion and dilaton fields.
The discretization of the axion's flux
implies the
quantization of the square of the
magnetic parameter.
However, the homology of spacetime does not predict a quantization of the U(1)
parameters. So it is surprising that a quantization condition for the
magnetic parameter is present. This result does not follow from topology
but rather from compactification in string theory.

Alternatively, we might focus on the solutions presented here as
an Einstein--Maxwell dilaton--axion configuration.
It is known that the only global charges of spacetime will be the
Mass and the U(1) charge and what we might call the
``dilaton charge'' is a combination of these global charges.
Furthermore, in the black hole case ($l=0$), a
Witten effect is observed \cite{witten}. The distinguished solutions, 
at $\Theta=0,\pi/2$, lack an axion field and
so are absent of a local $\theta$ term.
Upon an SL(2,$\mathbb{R}$) rotation through an angle
$\Theta$, they acquire and axion flux and at the same
time a dyonic status in the Maxwell sector.

The family of solutions presented here originate from applying an O(1,1)
boost and then an SL(2,$\mathbb{R}$) transformation. These
two operations have been shown  to not commute \cite{bmq}.
This means that when further transformations are applied, new solutions can be found.
This method allows for a six-parameter dilatonic Taub--NUT solution whose independent
physical quantities are electric and magnetic mass and charge (Maxwell and scalar).
In this most general setting, the axion's flux is still quantized. It is the fundamental
result presented here which transcends to this larger family. Since the axionic parameter
is independent from the rest, no discretization is present in this larger class for any of the
other parameters. The Kalb--Ramond field in this class of solutions is in general twisted,
but always untwists at the asymptotic boundary, as discussed in Section \ref{axidil}.
The class also specializes to a subclass with a zero valued Maxwell field.
This subclass of dilatonic Taub--NUT spaces has Kalb--Ramond fields classified by ordinary cohomology.
The general class of Ref.\ \cite{bmq} contains in the static limit the Schwarzschild black hole
and other static dilatonic spacetimes. However, the latter are singular on the horizon.
The dilatonic Taub--NUT spacetimes of this family are all nonsingular, but we argue that
the most physically interesting solutions are those of \cite{jm,ss}. They are, after all,
the closest NUT space to the static dilatonic black hole.

Let us also note that quantized axion--dilaton
gravitational configurations have been previously 
considered in the literature, see, e.g., \cite{kt}.
There, charge quantization and duality symmetry are
studied for axion--dilaton black holes of zero entropy and zero horizon area.
This stable particle-like class has charge quantizations
which are not consistent with the full SL(2,$\mathbb{R}$) transformations 
and is broken into SL(2,$\mathbb{Z}$) or one of its subgroups.
The asymptotic values of the scalar field pair are
associated to the modular parameter of a complex torus.
As a consequence electric/magnetic charge quantization is
related to winding numbers.
Indeed, our results effectively share common features with those of \cite{kt}.

\section*{Acknowledgements}
DFA would like to thank Clifford Johnson and Leonardo Pati\~no 
for useful conversations.
He also thanks the members and staff of the Department of
Physics and Astronomy at
the University of Southern California, where part of this work was done,
for their hospitality
and the Theoretical High Energy Physics Group for a stimulating atmosphere.
We thank the anonymous referees for their critical and helpful comments. 
DFA was supported by CONACyT under Grant No. 404449 and by mobility grants 
from PAEP-UNAM and CONACyT.
This work has been supported by the UNAM-DGAPA-PAPIIT, Grant No. IN111617.

\end{document}